\documentclass[journal=nalefd,manuscript=article,layout=onecolumn]{achemso}
\usepackage[version=3]{mhchem} % Formula subscripts using \ce{}
\usepackage[T1]{fontenc}
\usepackage{amsmath}  
\usepackage{color}     % Use modern font encodings
 \usepackage{hyperref}
\author{Mickael Buret}
\affiliation[Universit\'e de Bourgogne]{Laboratoire Interdisciplinaire Carnot de Bourgogne UMR 6303, CNRS-Universit\'e de Bourgogne Franche-Comt\'e, 21078 Dijon, France}

\author{Alexander V. Uskov}
\altaffiliation{ITMO University, Kronverkskiy 49, 197101, St. Petersburg, Russia}
\affiliation[Lebedev Physical Institute]
{Lebedev Physical Institute, Moscow, Russia}

\author{Jean Dellinger}
\altaffiliation{T\'el\'ecom Physique Strasbourg 67412 Illkirch, France}
\affiliation[Universit\'e de Bourgogne]{Laboratoire Interdisciplinaire Carnot de Bourgogne UMR 6303, CNRS-Universit\'e de Bourgogne Franche-Comt\'e, 21078 Dijon, France}

\author{Nicolas Cazier}
\affiliation[Universit\'e de Bourgogne]{Laboratoire Interdisciplinaire Carnot de Bourgogne UMR 6303, CNRS-Universit\'e de Bourgogne Franche-Comt\'e, 21078 Dijon, France}

\author{Marie-Maxime Mennemanteuil}
\affiliation[Universit\'e de Bourgogne]{Laboratoire Interdisciplinaire Carnot de Bourgogne UMR 6303, CNRS-Universit\'e de Bourgogne Franche-Comt\'e, 21078 Dijon, France}

\author{Johann Berthelot}
\altaffiliation{The Institute of Photonic
Sciences, 08860 Castelldefels, Spain}
\affiliation[Universit\'e de Bourgogne]{Laboratoire Interdisciplinaire Carnot de Bourgogne UMR 6303, CNRS-Universit\'e de Bourgogne Franche-Comt\'e, 21078 Dijon, France}

\author{Igor V. Smetanin}
\affiliation[Lebedev Physical Institute]
{Lebedev Physical Institute, Moscow, Russia}

\author{Igor E. Protsenko}
\affiliation[Lebedev Physical Institute]
{Lebedev Physical Institute, Moscow, Russia}

\author{G\'erard Colas-des-Francs}
\affiliation[Universit\'e de Bourgogne]{Laboratoire Interdisciplinaire Carnot de Bourgogne UMR 6303, CNRS-Universit\'e de Bourgogne Franche-Comt\'e, 21078 Dijon, France}

\author{Alexandre Bouhelier}
\affiliation[Universit\'e de Bourgogne]{Laboratoire Interdisciplinaire Carnot de Bourgogne UMR 6303, CNRS-Universit\'e de Bourgogne Franche-Comt\'e, 21078 Dijon, France}
\email{alexandre.bouhelier@u-bourgogne.fr}
\phone{+33 38039 60222}
\fax{+33 38039 6024}

\title[Electron-fed optical antenna]
  {Spontaneous hot-electron light emission from electron-fed optical antennas}


\begin{document}

%\begin{tocentry}

%\includegraphics[height=3.5cm]{TOC}

%\end{tocentry}

\begin{abstract}

Nanoscale electronics and photonics are among the most promising research areas providing functional nano-components for data transfer and signal processing. By adopting metal-based optical antennas as a disruptive technological vehicle, we demonstrate that these two device-generating technologies can be interfaced to create an electronically-driven self-emitting unit. This nanoscale plasmonic transmitter operates by injecting electrons in a contacted tunneling antenna feedgap. Under certain operating conditions, we show that the antenna enters a highly nonlinear regime in which the energy of the emitted photons exceeds the quantum limit imposed by the applied bias. We propose a model based upon the spontaneous emission of hot electrons that correctly reproduces the experimental findings. The electron-fed optical antennas described here are critical devices for interfacing electrons and photons, enabling thus the development of optical transceivers for on-chip wireless broadcasting of information at the nanoscale. 

\end{abstract}


Optical antennas are designed arrangements of metal nanoparticles operating at the surface plasmon resonance. These nanoscale devices are passive wave-vector converters largely used for electromagnetic interfacing and radiation engineering~\cite{novotny09antenna,novotny11NP}. Optical antennas are pervasive in a growing number of disciplines including single-emitter control~\cite{curto10}, high-harmonic generation~\cite{kimNature08},  or hot-carriers production~\cite{brongersma15}. In these diverse applications, an external light field drives the antenna which acts as a relaying element between an in-coupling optical stimuli and the desired out-coupled response. However, an appealing feature of metal-based plasmonic units is their ability to process optical signals and electric currents via a shared circuitry. This unique asset recently fostered the development of planar electrically-activated surface plasmon polariton sources thereby addressing the long-standing issue of on-chip integration~\cite{vandorpe10,polman10,brongersma12NL,RaiPRL13}. In this paper, we pursue this incentive and demonstrate the conversion of an electrical power to an electromagnetic radiation inside the feedgap of an optical antenna. This nanoscale transducing element is an essential component for interfacing an electronic layer with a photon-based platform, and may enable a wireless broadcasting link~\cite{engheta10} when paired with matching optical rectennas~\cite{ward10,Stolz2014}. 

Our approach is based on electrically pumping the feedback region of a tunneling optical gap antenna~\cite{HechtNL12,Stolz2014}. Upon injecting electrons, we record a highly nonlinear energy-forbidden light emission from the feedgap.  We show that this unconventional radiation is linked to the temperature of the electron sub-system and the underlying surface plasmon resonances. By appropriately positioning the feedgap with respect to the leads, we demonstrate an agility of the angular distribution of the emitted photons with an increase of the directivity. 

\begin{figure}
\includegraphics[width=9cm]{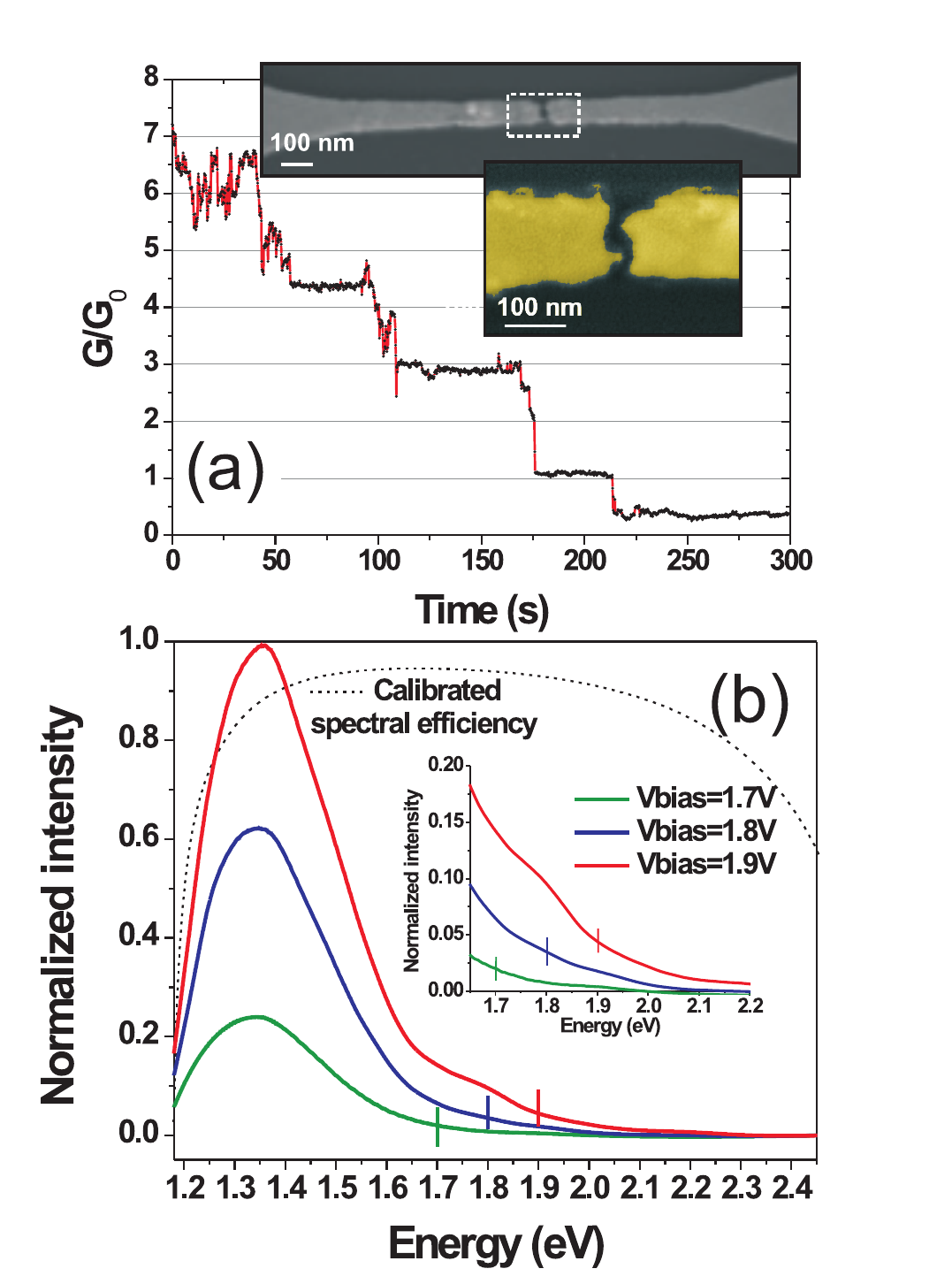}
    \caption{(a) Evolution of the normalized conductance during the last moments of the electromigration process.  Quantum conductance steps are clearly marked. Inset: scanning electron micrograph of a Au nanowire after electromigration and close-up view of the junction area. (b) Emission spectra showing the displacement of the high energy side with the applied bias. The spectra are not corrected for the detection efficiency (dash curve). The vertical bars indicate the quantum limit $h\nu_{\rm max} = eV_{\rm bias}$. Inset: magnified energy distributions near the quantum limit showing over-bias photon energy.}
  \label{quantumsteps}
\end{figure}

In-plane tunneling optical gap antennas are realized by a controlled electromigration of a 100~nm wide 4~$\mu$m long Au nanowire. The electromigration and subsequent optical and electrical characterizations discussed below are perfomed under ambient conditions using the apparatus sketched in the supplementary file. The nanowire and the macroscopic electrodes are deposited on a glass coverslip by a double-step lithography involving electron-beam writing followed by an ultraviolet patterning. A 2~nm Cr layer is  thermally evaporated to favor the adhesion of a 50~nm thick Au layer. A liftoff of the resist finalizes the structure.  Electromigration of the nanowire is obtained by constantly monitoring the time evolution of the nanowire conductance $G$ upon applying a slowly increasing voltage  $V_{\rm bias}$. When the conductance drops below a predetermined threshold due to Joule heating and the onset of electromigration,  $V_{\rm bias}$ is slightly reduced to contain the variation of $G$. This manually operated feedback of the bias voltage is maintained during the complete electrical thinning of the nanowire. Figure~\ref{quantumsteps}(a) displays the last minutes of the process where steps in units of the quantum conductance $G_0=2e^2/h$ are clearly observed indicating the passage from a ballistic electron transport to the tunneling regime when $G<G_0$ ($t>$220~s). Here $e$ is the charge of an electron and $h$ is the Planck's constant. The inset of Fig.~\ref{quantumsteps}(a) shows a scanning electron micrograph of a typical electromigrated nanowire together with a close-up view of the junction area after being used. Note that a conducting layer of Au was evaporated on the sample to enable SEM imaging. The junction separating the two electrodes acts as a non-resonant optical gap antenna where the optical response and the electrical potential are self-aligned in a nanometer-scale feedgap~\cite{ward10,Berthelot:12OPEX,Stolz2014}. 

Light emission observed from biased tunnel junctions is well documented in the literature since the pioneering work of Lambe and McCarthy~\cite{lamb76}. Photon emission is generally understood as the radiative decay of surface plasmon modes. These modes can be either excited within the junction by inelastic tunneling current fluctuations~\cite{sparks89,apell90}, or directly in the metal electrodes by hot carriers relaxation~\cite{Dimaria83,Despax93}.  A particularity of the emission spectrum is that the high-energy side is bound by the quantum limit where the quantum of energy carried by the photons cannot exceed the electron energy provided by the bias voltage: $h\nu_{\rm max} \leq eV_{\rm bias}$ where $\nu_{\rm max}$ is the highest frequency component of the spectrum. We do observe such bias-controlled spectral distribution in electromigrated optical gap antennas characterized by relatively low zero-bias conductance $G\simeq 10^{-2}G_0$ and smaller. A representative example is illustrated in Fig.~\ref{quantumsteps}(b) for three operation voltages of the light-emitting antenna. As expected from a tunneling process, increasing $V_{\rm bias}$ from 1.7~V to 1.9~V leads to a higher number of charges injected in the antenna feedgap and we consequently observe a net gain of the photon flux. The vertical bars represent the position of the quantum limit $h\nu_{\rm max} = eV_{\rm bias}$ limiting the energy of the released photons. A surface plasmon contribution at 1.8~eV can be seen as a shoulder in the spectrum taken at $V_{\rm bias}$=1.9~V. This spectral feature is not observed for smaller biases because the electron energy is not sufficient to populate the plasmon mode. This overall behavior has recently been reported and exploited in resonant systems by J. Kern \emph{et al.} ~\cite{Hecht15} and will not be discussed further here. 

\begin{figure}
\includegraphics[width=9cm]{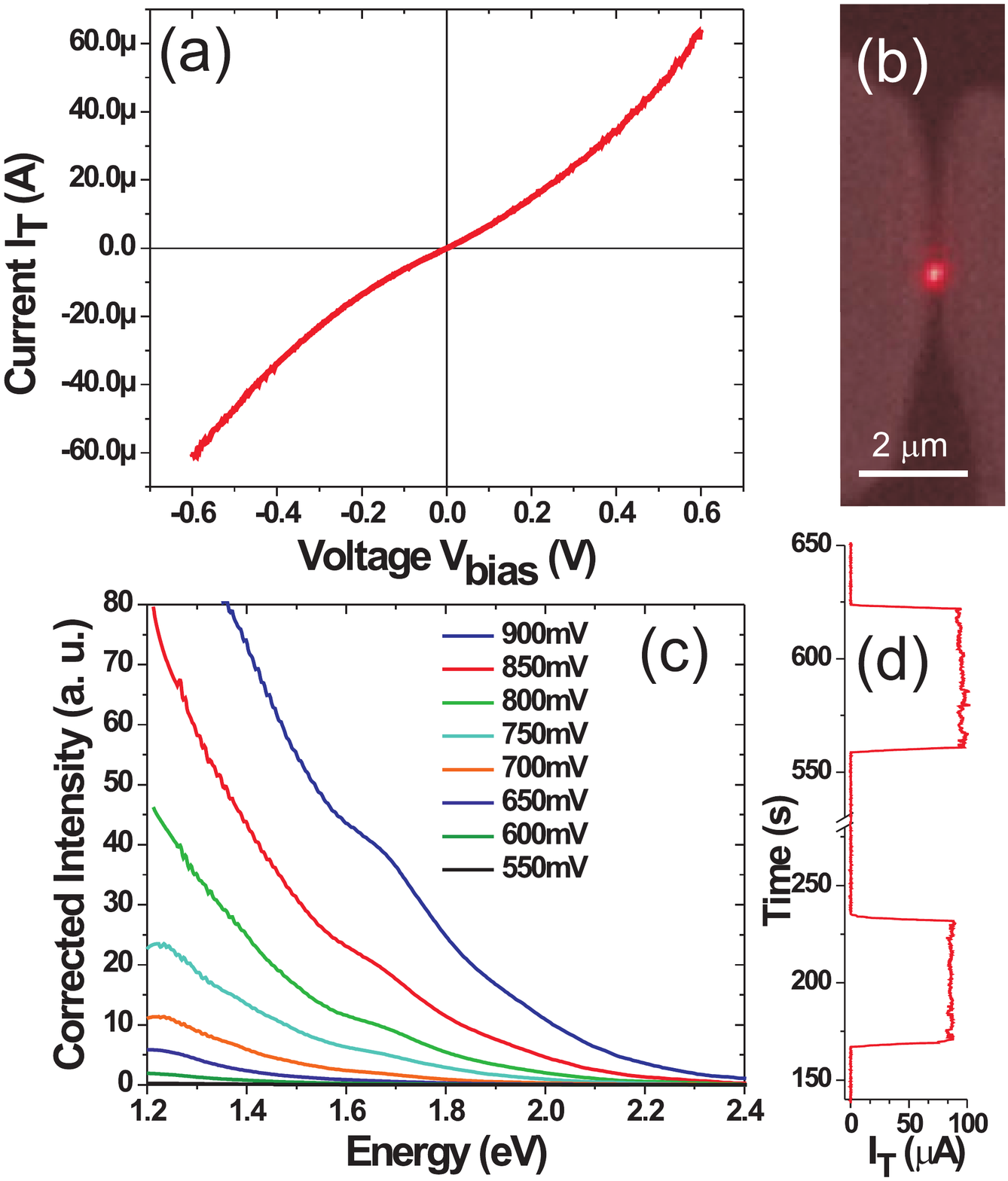}
    \caption{(a) Output characteristics of the $G \sim 0.8G_0$ junction  featuring a large tunnel current. (b) Wide-field optical image in false color of the electron-fed optical antenna operated at $V_{\rm bias}$=1V. A residual illumination enables a visualization of the contacting electrodes (darker areas).  (c) Emission spectra of the device for different bias voltages. The emission covers much of the visible spectral domain in clear deviation from the quantum cutoff imposing $h\nu_{\rm max} \leq eV_{\rm bias}$. All spectra are corrected by the calibrated efficiency curve displayed in Fig.~\ref{quantumsteps}(b). (d) Time trace showing the stability of the tunneling current $I_T$ during two 70~s sequential acquisitions at $V_{\rm bias}$= 750~mV and 800~mV, respectively.  Time bin is 10~ms.}
  \label{smalljunction}
\end{figure}

With a careful observation of the emission spectra, it is conspicuous that they are not completely bound by the quantum limit. We record a small portion of the spectral distributions clearly violating the $h\nu_{\rm max} = eV_{\rm bias}$ cutoff as indicated by the inset of Fig.~\ref{quantumsteps}(b) showing a close-up view of the spectra near the three thresholds. This unconventional light is even more pronounced for optical tunneling gap antenna with high zero-bias conductance. Figure~\ref{smalljunction}(a) shows  
the nonlinear output characteristic relating the tunneling current $I_T$ to the bias for an electromigrated junction with $G\sim 0.8G_0$. Noticeable is the unusual large current $I_T$ tunneling through the feedgap for moderate biases compared to the nA range that is typical for larger tunneling gaps. The current-to-voltage characteristics does not show any signature of molecular adsorbate in the junction. All the measurement made in the following were acquired from freshly electromigrated nanowires mitigating thus the possible adsorption of contaminants in the junction.  Upon injecting charges, we observe an optical response from the antenna feedgap as illustrated by the wide-field optical micrograph of Fig.~\ref{smalljunction}(b). In this image, a weak diascopic illumination enables the visualization of the device geometry, and in particular the layout of the contacting electrodes. The lateral dimension of the emission spot is limited by the resolving power of the objective ($\times$100;  N. A.=1.49). The spectral characteristics of this device functioning at ambient conditions are reported in Fig.~\ref{smalljunction}(c) for different voltages $V_{\rm bias}$ comprised between 550~mV and 900~mV. The emission manifestly covers a large portion of the visible spectrum with  $h\nu_{\rm max} > eV_{\rm bias}$ in contrast with Fig.~\ref{quantumsteps}(b). This energy distribution is at clear variance from the quantum limit, and cannot be accounted for by the standard picture of an inelastic single electron process. 
Figure~\ref{smalljunction}(d) shows the electrical stability of the junction to within a few percent during two consecutive 70~s acquisitions of the spectra for $V_{\rm bias}$= 750~mV and 800~mV. 

Overbias light emission in Au junctions has been occasionally reported in the context of scanning tunneling microscopy (STM)~\cite{pechou98,welland02,Schull09,Schneider13}. The radiation mechanism was first described as a spontaneous emission from an elevated temperature of the electron system~\cite{welland02}. This interpretation was then discarded on account for the presence of characteristics plasmon modes in the spectra~\cite{Schull09}. An Auger-like process of hot carriers was introduced to increase the electronic energy distribution responsible for the $h\nu_{\rm max} > eV_{\rm bias}$ emission~\cite{Schull09,Schneider13}. A quantitative agreement with the experiment was obtained using the framework of dynamical Coulomb blockade theory for processes implying  the coherent interaction of two electrons~\cite{belzig14}. 

\begin{figure}
\includegraphics[width=9cm]{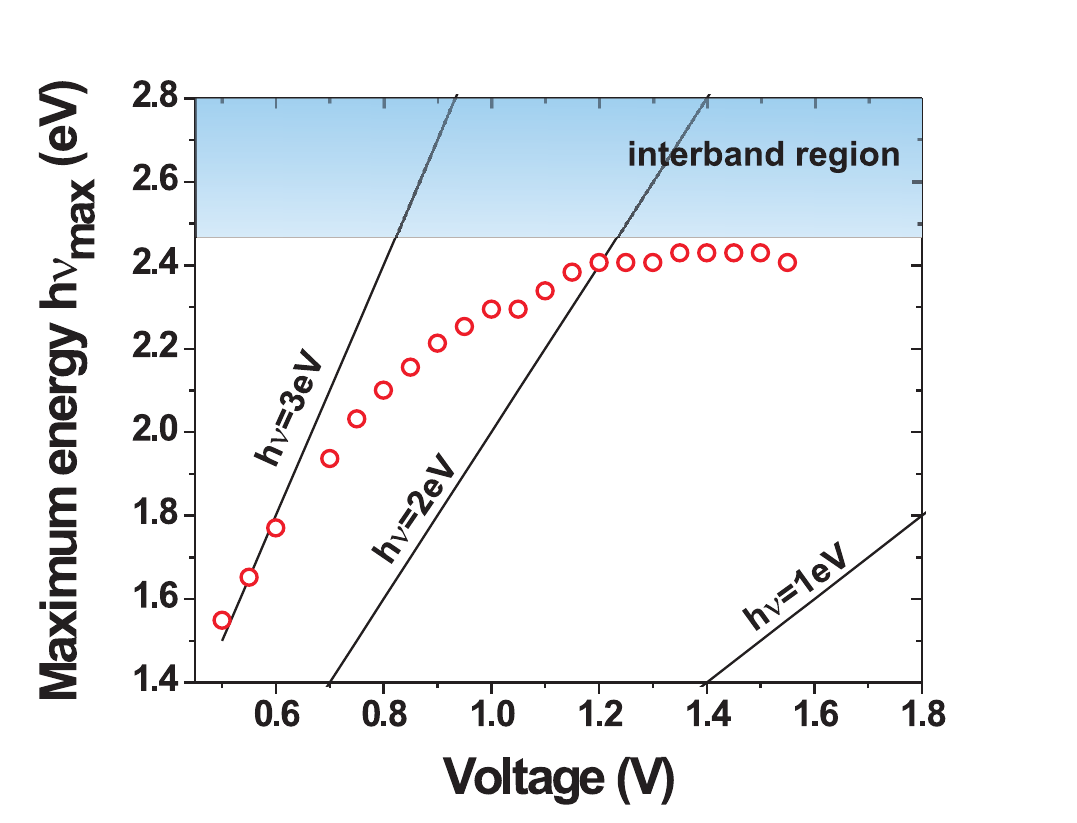}
    \caption{Evolution of the highest photon energy with voltage. The lines represent the energy conservation for a one-electron, a two-electron and a three-electron processes, respectively. The shaded area is the energy region where interband transitions to $d$-band electrons dominate.}
  \label{energymax}
\end{figure}

Our results departs from a correlated two-electron process as illustrated in Fig.~\ref{energymax} showing the bias evolution of the largest photon energy $h\nu_{\rm max} $ emitted by the antenna.  For moderate biases, we record a photon energy requiring the contribution of three electrons. As the voltage increases,  $h\nu_{\rm max} $ shifts to higher energies following a saturation curve up to the asymptotic value of $\sim$2.4~eV. The saturation of $h\nu_{\rm max} $ strongly suggests an inhibition of the antenna emission due to the onset of Au interband reabsorption by low-lying $d$-band electrons~\cite{johnson72}. 

The quantized conduction step observed at $G/G_0$=1 demonstrate the formation of a single conduction channel between two Au atoms [Fig.~\ref{quantumsteps}(a)]. Because, the Fermi levels between the two sides of the feedgap are separated by $eV_{\rm bias}$, electrons tunneling through the antenna feedgap produce a hot carrier distribution in the receiving electrode. In a region near the contact, the injected power $P=I_TV_{\rm bias}$ is dissipated to the cold electrons of the drain electrode and causes a raise of their effective electronic temperature $T_e$. The electron temperature under steady-state current pumping can be found from a balance between the electrical power $P$ dissipated near the contact, and the cooling rate of the electrons in this region. Typically, the cooling of electrons in metal occurs though the electronic heat conductivity and the interaction of the electrons with bulk phonons~\cite{Gamaly11}.  However, the electron-bulk phonon interaction can be strongly suppressed in nanostructures with characteristic size $L$  shorter than the length $l_{\rm energy}$  characterizing the exchange of energy between electrons and bulk phonons~\cite{tomchuk90,Federovich00}: $L<l_{\rm energy}\sim v_{\rm F}\times \tau_{\rm energy}$, where $\tau_{\rm energy}$ is the characteristic time for electrons to exchange their energy with phonons and $v_{\rm F}$ is the Fermi velocity. Note that $l_{\rm energy}$ is different from the electron mean-free path $l_{e \rightarrow ph}\sim50-60$~nm~\cite{Chopra63,welland02} for electron-phonon elastic scattering. If $L<l_{\rm energy}$, the collision of electrons with the walls of the nanostructure becomes an important cooling mechanism with which electrons can loose their energy~\cite{Groeneveld95,tomchuk90,Federovich00}.

Tomchuk and Fedorovich developed a model~\cite{Tomchuk66} for electron cooling in a nanoparticle of the size $L$  in which electrons collide with the nanoparticle's surface with the frequency $v_{\rm F}/L$, and obtained a formula relating the electron temperature in the nanoparticle to the electrical power injected into the nanostructure:
\begin{equation}
(k_{\rm B}T_e)^2-(k_{\rm B}T_L)^2=\alpha I_TV_{\rm bias},
\label{TF}
\end{equation}
where $T_L$ is the lattice temperature, $k_B$ is the Boltzmann's constant and 
\begin{equation}
\alpha=\left(\frac{\pi^2}{4}\frac{(mL)^2}{M\hbar^3}E_{\rm F}\right)^{-1}.
\label{alpha}
\end{equation}
Here, $m$ and $M$ are the electron and atomic masses, $E_{\rm F}$ is the Fermi energy and $\hbar=h/2\pi$. In the strong heating regime  ($T_e>>T_L$), Eq.~\ref{TF} reduces to 

\begin{equation}
k_{\rm B}T_e \sim (\alpha I_TV_{\rm bias})^{1/2}.
\label{Te}
\end{equation}

Tomchuck and Fedorovich obtained Eq.~\ref{TF} to Eq.~\ref{Te} by neglecting the electronic heat conductivity as a mechanism for electron cooling. In the nanowire-like system forming the optical antenna discussed here, the electronic heat conductivity can however be an important aspect to consider~\cite{Natelson14}. Therefore, we developed a model described in the Supplementary Material in which we took into account cooling of the electron sub-system both by electron collision with the wall of the feedgap (modified from Tomchuck-Fedorovich model) and the electronic heat conductivity along nanowire. We  obtained formulas of the forms of Eq.~\ref{TF} to Eq.~\ref{Te} but with a different coefficient $\alpha=\alpha'$:
\begin{equation}
\alpha'=\frac{4k_B}{L^{3/2}m\pi\sqrt{2b\frac{E_F}{M\hbar^3}}}. 
\label{alpha_nw}
\end{equation}
Where $L$ is the mean-free path for electrons to collide with the surface. This distance depends on the ill-defined junction geometry resulting from the electromigration process and is thus proportional to the square root of an effective area $L\sim\sqrt{A_{\rm eff}}$.  $b$ is the electronic heat conductivity coefficient entering the thermal conductivity. For bulk materials, the thermal conductivity is expressed as 

\begin{equation}
\kappa=C_e v_{\rm F} \frac{l_{e \rightarrow ph}}{3} =\gamma T_e v_{\rm F} \frac{l_{e \rightarrow ph}}{3}=bT_e
\label{kappa}
\end{equation} 
where $C_e$ is the electronic heat capacity, $\gamma=\pi^2Nk_{\rm B}^2/2E_F$ is the Sommerfeld constant and $b=\gamma  v_{\rm F} l_{e \rightarrow ph}/3$. When the length of a system becomes comparable to $ l_{e \rightarrow ph}$, the thermophysical properties are affected by  scattering of electrons at surfaces and $b$ becomes size-dependent and can be substantially reduced for small nanowires~\cite{kelly08}.

An important feature of the cooling mechanism described by Eq.\ref{TF} to Eq.~\ref{alpha_nw} is that the electron temperature  $T_e$ is proportional to square root of the electrical power  $P=I_TV_{\rm bias}$. In contrast, when electron cooling occurs by exchanging their energy to bulk phonons, the electron temperature is not proportional to square root the power $P$. For instance, if the electron heat conductivity is neglected and cooling is defined only by electron-bulk phonon interaction, the electron temperature is set by the electrical power injected in the system $T_e\propto P$.  Thus, the square root dependence can be considered as confirmation of an electron cooling through surface collisions rather than electron scattering to bulk phonons.

Electron characterized by a temperature $T_e$ spontaneously radiate with an emission spectrum $U(\nu)$ given by 
\begin{equation}
U(\nu,\textbf{r})= \rho(\nu,\textbf{r}) \frac{h\nu}{\exp(h\nu/k_{\rm B}T_e)-1}.
\label{planck}
\end{equation}
$\rho(\nu,\textbf{r})$ is the density of modes with a frequency $\nu$ into which the emission occurs~\cite{joulain03}. The exponent $\exp(h\nu/k_{\rm B}T)$ originates from the population of the electronic energy levels participating to the emission. 
With the condition $\exp(h\nu/k_{\rm B}T_e)>>1$, Eq.~\ref{planck} can then be rewritten in the form:
\begin{equation}
{\rm ln}[U(\nu,\textbf{r})]= {\rm ln}[\rho(\nu,\textbf{r})h\nu]-\frac{h\nu}{k_{\rm B}T_e}.
\label{plancklinear}
\end{equation}

\begin{figure}
\includegraphics[width=9cm]{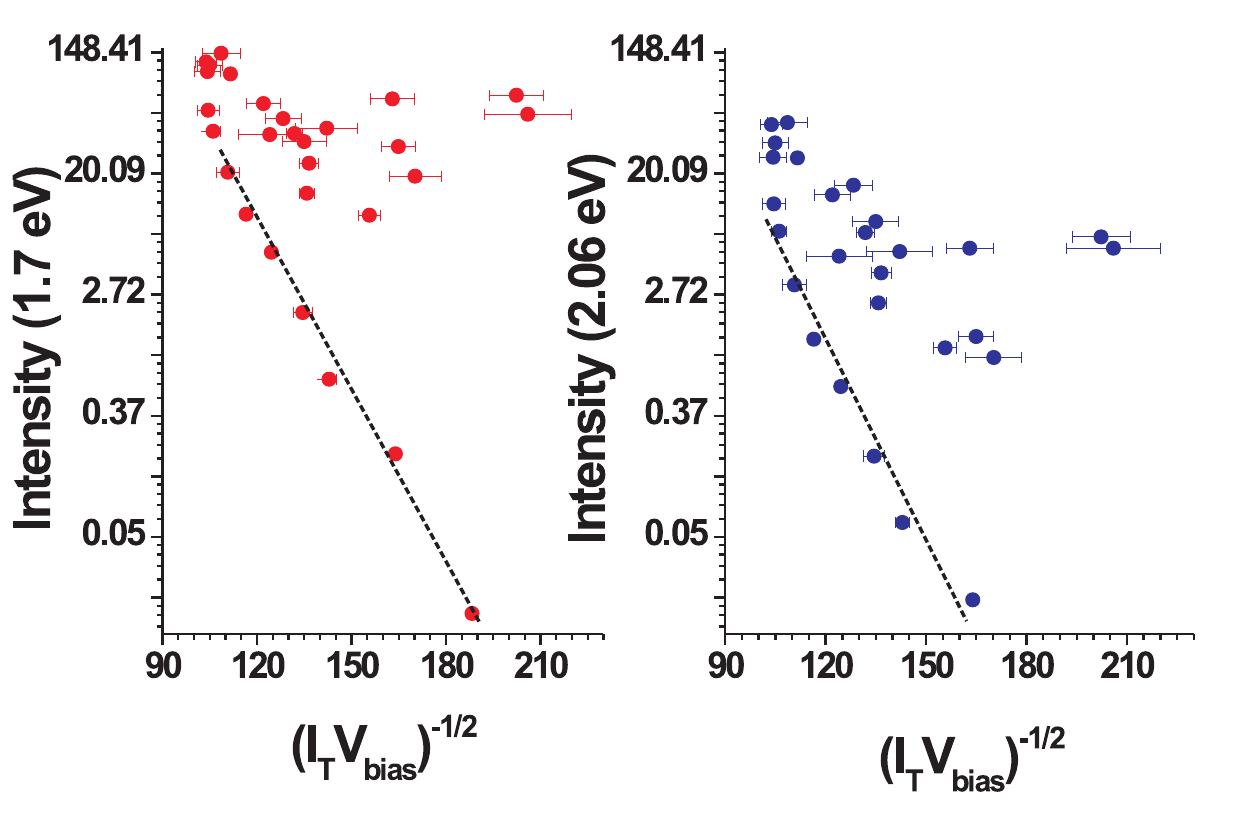}
    \caption{Semi-logarithmic plot of the light intensity extracted at 1.7~eV and 2.06~eV showing a clear linear dependence with the square root of the electrical power. The data points outside this trend correspond to an unstable electrical operation of the antenna (degradation). The current fluctuations occurring during the spectral acquisitions are accounted for by the error bars.}
  \label{linear}
\end{figure}

Feeding Eq.~\ref{Te} in Eq.~\ref{plancklinear}, the intensity of the light at a given frequency $\nu$  spontaneously emitted from the hot electron distribution should scale linearly with $(I_TV_{\rm bias})^{-1/2}$ in a semi-logarithmic plot~\cite{welland02}.  Figure~\ref{linear} shows semi-logarithmic plots of the light intensity extracted from the spectra at 1.7~eV and 2.06~eV as a function of the variable $(I_TV_{\rm bias})^{-1/2}$, respectively. There is clear linear dependence (dash lines) measured when the junction operates in stable conditions. For data points outside of this trend, the tunneling current $I_T$ is erratic with  $V_{\rm bias}$ suggesting a degradation of the feedgap through the modification of the junction's conductance. Importantly, and this is the main argument of the paper, the points aligned along the dashed lines are confirming the hot electron origin of the antenna emission, in line with early experiments in island metal films~\cite{Federovich00} and STM~\cite{welland02}. When the hot electrons collide with the the nanoantenna surface they radiate via the available modes of the structure by a Bremsstrahlung process~\cite{Baratoff92} and a Cerenkov-like radiation~\cite{garciadeabajo10,Khurgin14} to create a thermal distribution at quasi-equilibrium. 

From the slope of the linear fits, the parameter $\alpha$ entering Eq.~\ref{Te} can be readily inferred and an estimation of the effective electron temperature $T_e$ can be made using Eq.~\ref{Te}. Figure~\ref{ElectronT}(a) shows the evolution of $T_e$ with the tunneling current $I_T$. Under such elevated current passing the tunnel junction, the electron temperature can reach values corresponding to electron energies 80~meV to 170~meV above the Fermi level. These temperatures are comparable to those reported for nanoparticles excited with ultra-short optical laser pulses with intensities below the damage threshold~\cite{Gamaly11}. Increasing further the electrical power injected in the hot electron gas does not necessarily rise the intensity of the antenna glow as shown in Fig.~\ref{ElectronT}(b) for photons emitted at 1.7~eV.  The electromagnetic energy emitted by the hot electrons steeply rises until $T_e$=2000~K. After this electron temperature, the curve inflects and the nonlinearity reduces. The solid line is a fit to the data using Eq.~\ref{planck} leaving the density of states $\rho$ as free parameter. The electron temperature inferred using Eq.~\ref{Te} to Eq.~\ref{alpha_nw} indicates that with the electrical conditions experimentally used here, the antenna already operates in its highest nonlinear regime.  Despite the strong nonlinearity of the process [Fig.~\ref{ElectronT}(b)], at $T_e$=2000~K the estimated external conversion yield remains low at $\sim10^{-11}$ photon/electron {reflecting the limited spectral coverage of the detection.

\begin{figure}
\includegraphics[width=9cm]{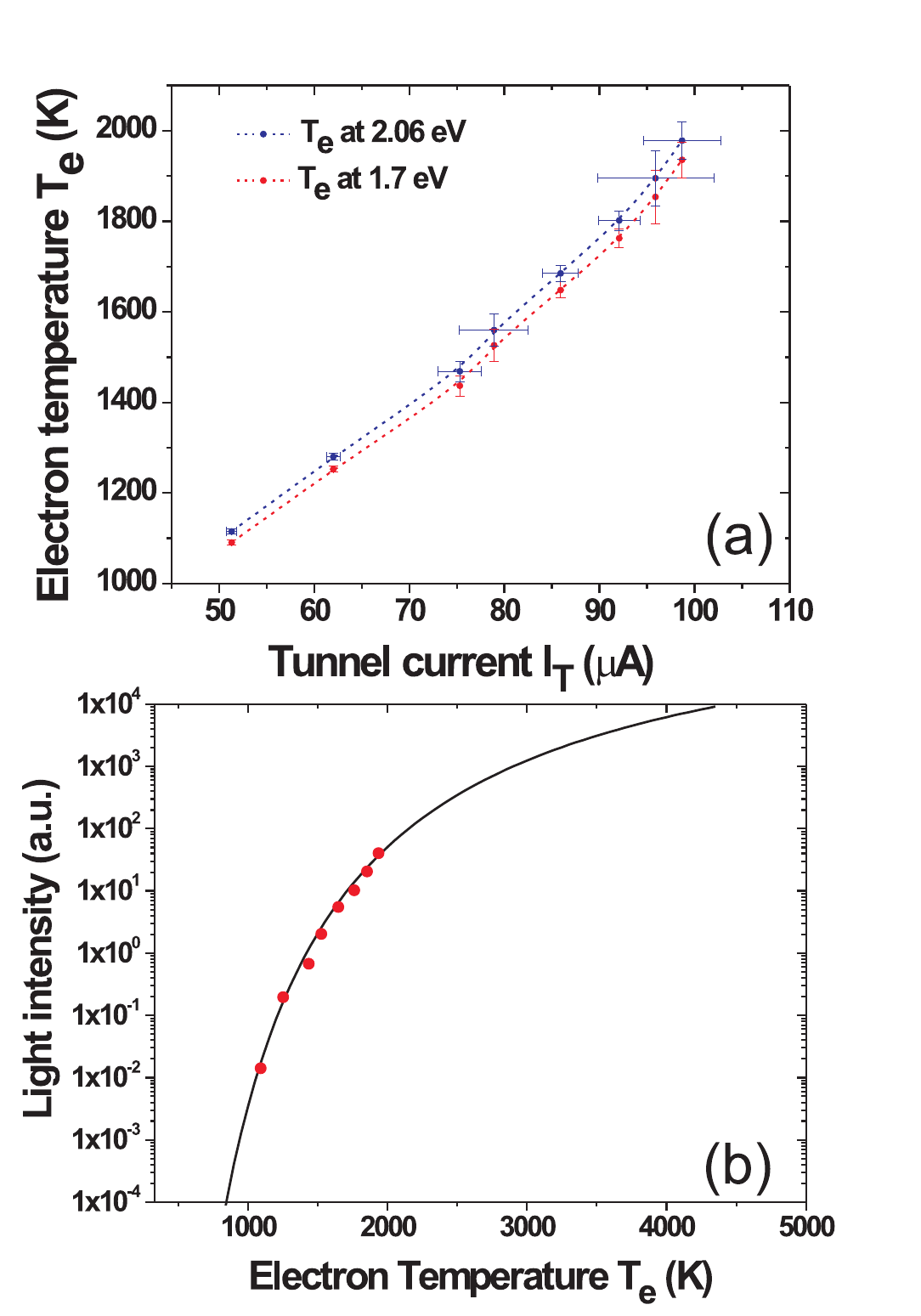}
    \caption{(a) Estimated effective electron temperature $T_e$ as a function of the tunnel current for the two energies illustrated in Fig.\ref{linear}. (b) Light intensity versus electron temperature (semi-logarithmic scale). The red points are the inferred electron temperature and the solid line is the evolution of the light intensity at 1.7~eV predicted by Eq.~\ref{planck}.}
  \label{ElectronT}
\end{figure}

Using Eq.\ref{TF} to Eq.~\ref{alpha_nw}, we estimate the upper and lower bounds for the characteristic length describing electron collisions with surface. To do this, we used the parameter $\alpha$ deduced from the linear fits in Fig.~\ref{linear} and feed it to Eq.~\ref{alpha_nw} together with the reported values for the size-dependence of the thermal conductivity~\cite{kelly08}.  We find 13~nm$<L<$33~nm for $b$ corresponding to bulk and $b$ estimated from a 1~nm thick nanowire.  We see that this range is less than the electron cooling length $l_{\rm energy}$ and the electron mean free path $l_{e \rightarrow ph}$ \emph{i.e}.,  $L<l_{\rm energy},l_{e \rightarrow ph}$  which is the necessary condition for applying the model. This value is somewhat consistent with the effective area of the tunnel junction made by electromigration (see inset of Fig.~\ref{quantumsteps}).  Thus, the application of the adapted Tomchuk and Fedorovich's model with the inclusion of electron heat conductivity  to describe electron heating and cooling in tunneling optical gap antennas discussed here seems to be self-consistent and justified.

For comparison purposes, we would like to discuss our results in the light of a recent contribution where the electron temperature in a ballistic nano-constriction was estimated from noise measurement~\cite{NatelsonSR14}. The electronic heating in the constriction was assigned to the viscosity of the quantum electronic fluid~\cite{DiVentra06,DiVentra09,DiVentra11} and the cooling was insured by conventional electron heat conductivity. The authors found that about 2\% of the injected electrical power is dissipated in the constriction for junction's conductances in the range of a few $G_0$. In that contribution, the electron temperature in the constriction can again be written as Eq.~\ref{TF} and Eq.~\ref{Te} but with a different coefficient $\alpha=\alpha''$.  We compare  $\alpha''$ with $\alpha'$  given by Eq.~\ref{alpha_nw} in the Supplementary Material, and find that both models predict an electronic temperature proportional to the square root of the electrical power fed into the system. The difference of the proportions used to elevate the temperature of the electron subsystem between the models is consistent with $T_e$ inferred in both set of measurements. 

At that point, it is interesting to come back to role of surface plasmon in the emission spectra. The thermal radiation mechanism was ruled out in STM measurement because the plasmon modes recorded in the emission spectra were not consistent with a blackbody-like glow~\cite{Schull09,Schneider13}. However, from Eq.~\ref{planck}, the energy released by the hot electrons is spectrally affected by the local density of optical states $\rho (\nu)$ at the position of the antenna feed. The presence of plasmon modes contributes drastically to increase this quantity at their resonance energies~\cite{joulain03,Imura05b} and their spectral signatures are therefore expected in the emission spectra provided that the electron energy is sufficient to populate the plasmon states. Looking back at Fig.~\ref{smalljunction}(c), a clear shoulder becomes visible at 1.6--1.7~eV for biases of 750~mV and higher indicating the excitation of a plasmon mode. The strength of the plasmon is weak comparatively to the other part of the spectrum as expected from such small tunneling gap~\cite{aizpurua10CPT,Mortensen14}. Figure~\ref{plasmon}(a) reproduces the antenna's emission spectrum at $V_{\rm bias}$=850~mV. To confirm the hot electron origin of the emission, we fit the experimental spectrum by Eq.~\ref{planck}. A plasmon contribution is explicitly added in the density of states in the form of a Gaussian function $\rho_{\rm{SP}}(\nu)=\rho_0 \rm{exp}[-(h\nu_{\rm SP}-h\nu)/\sigma^2]$ where $\rho_0$, $h\nu_{\rm SP}$  and $\sigma$ are the plasmon's amplitude, resonance energy and spectral width, respectively. The black line in Fig.~\ref{plasmon}(a) shows the result of the fit for a plasmon centered at 1.63~eV with a spectral width of 0.12~eV and an electronic temperature $T_e$=1938~K.  The fit shows a reasonable agreement and the deduced electronic temperature confirms the value inferred in Fig.~\ref{ElectronT}(a). The residual of the fit indicates a discrepancy around 1.4~eV and 1.8~eV suggesting the presence of additional surface plasmon modes expected in this complex feedgap geometry~\cite{dawson09}. 

An important aspect concerns the polarization characteristics of the antenna emission. The thermal energy radiated from sub-wavelength wires was shown to contain a polarized component depending on the ratio between the radius of the homogeneously heated line-like structure and the emission wavelength~\cite{Hamann09}. However, for spatially-confined thermal source no net polarization was found~\cite{Natelson08}. For spectra of the type displayed in Fig.~\ref{plasmon}(a), the polarization state of the underlying plasmon modes is difficult to assess because the direction of the electric field will depend on the detected energy and symmetry of the modes~\cite{vanHulst13}. To mitigate the role of  surface plasmon resonances in the polarization response, we fabricated a tunneling optical gap antenna from two overlapping bowtie like electrodes (\emph{i.e.} without a metal bridge) and inserted an analyzer in front of the spectrograph. The emission spectra for different orientations of the analyzer are displayed in Fig.~\ref{plasmon}(b). The spectra essentially feature a black-body emission with no significant plasmonic contributions. Rotating the analyzer clearly shows that the emission from the feedgap is quasi-unpolarized as expected from a local thermal source. We measure a degree of linear polarization (DOP) of 0.11 at 1.3~ eV.

\begin{figure}
\includegraphics[width=9cm]{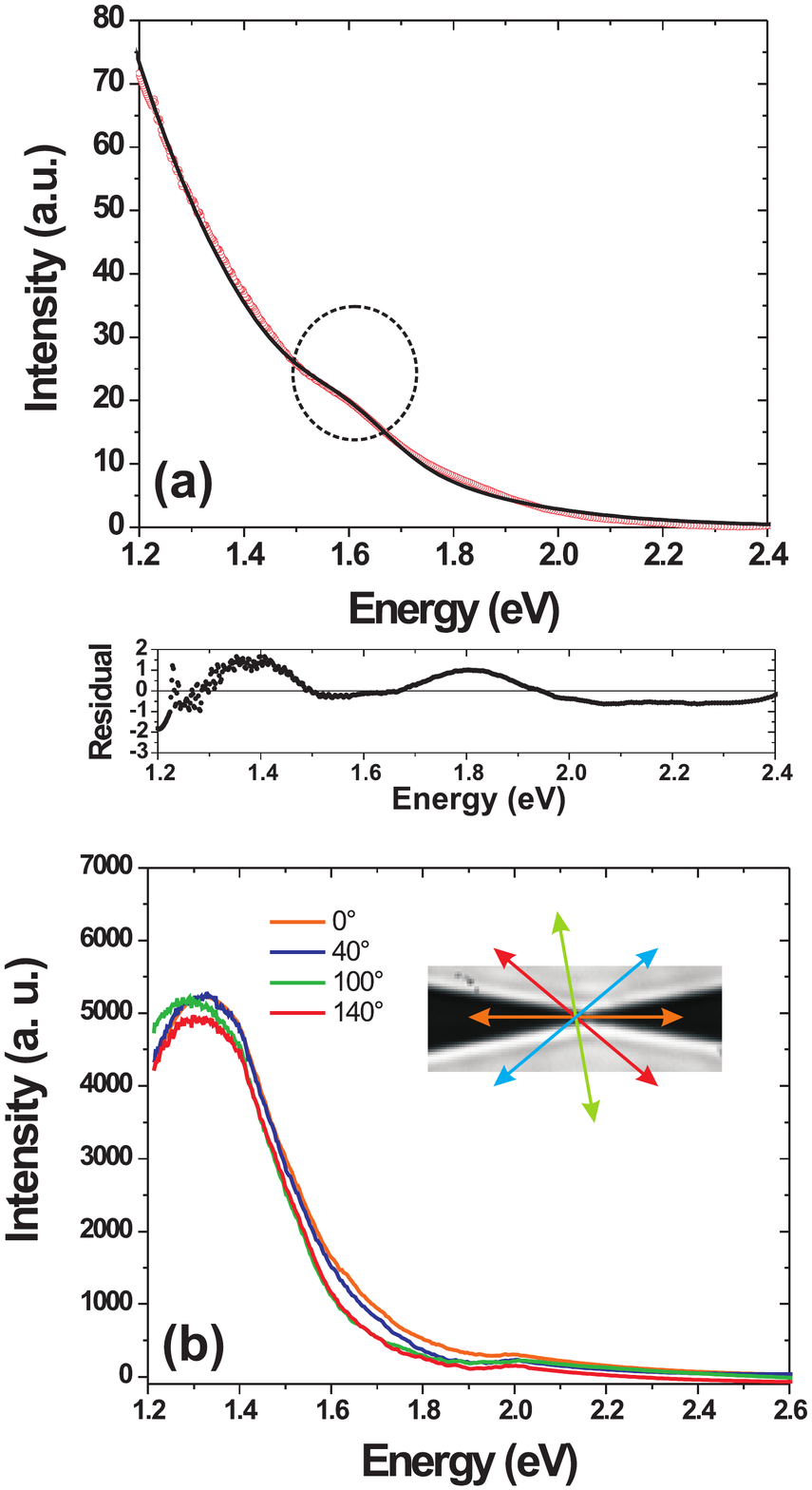}
 \caption{(a) Emission spectrum of the electron pumped antenna operated at 850~mV (red points) with a shoulder (circle) indicating the presence of a weak surface plasmon resonance. The black line is the expected spontaneous emission (Eq.~\ref{planck}) from an electron temperature bath of 1938~K and a local density of states featuring a single surface plasmon resonance at 1.63~eV. Inset: residues of the fit  suggesting the presence of additional plasmon modes at 1.8~eV and 1.4~eV. (b) Polarization response of an electron-fed antenna in absence of any significant plasmonic contributions.  $V_{\rm bias}$=1~V. The emission is quasi-unpolarized with a DOP of 0.11 at 1.3 eV. Inset: orientation of the analyzer overlaid to an optical image of the touching bowtie-like electrodes.}
  \label{plasmon}
\end{figure}

Finally, we investigate the emission diagram of the electron-pumped antennas. The emission pattern is an important characteristic of an optical antenna as it dictates the angular distribution of the released optical power. We directly measure the radiation diagram of the electrically-excited optical gap antennas by visualizing the emitted photons in the conjugate Fourier plane of the microscope~\cite{huang08prb} (see Supplementary Materials). A representative example of a self-emitting antenna is illustrated in Fig.~\ref{twolobes}(a). In this optical transmission image, the layout of the electrodes is readily seen together with a series of connected nanowires. A tunnel junction has been created on the nanowire indicated by the arrow and a diffraction-limited luminous spot is observed upon electrical biasing (see inset). Figure~\ref{twolobes}(b) shows the corresponding Fourier plane of the light emitted in the glass substrate. The diagram shows a strong emission located between the detection limit given by the numerical aperture of the objective and the critical angle at the glass/air interface. The radiation consists of two symetric lobes, which is however different to that of a dipole because the maxima are oriented along the nanowire axis. Although emitted locally, the radiation pattern is governed by the entire antenna geometry including the presence of the nanowire electrodes~\cite{Taminiau11,Brongersma13}, offering thus a certain degree of tunability. Since the tunneling gap is not perfectly centered at the middle of the leads and considering the large emission bandwidth, it is difficult to estimate the modal order of the emission. The directivity $D$, defined by the emitted power at the emission maximum normalized by the averaged radiated power~\cite{novotny09antenna}, is here measured at 16.5~dB. 

\begin{figure}
\includegraphics[width=9cm]{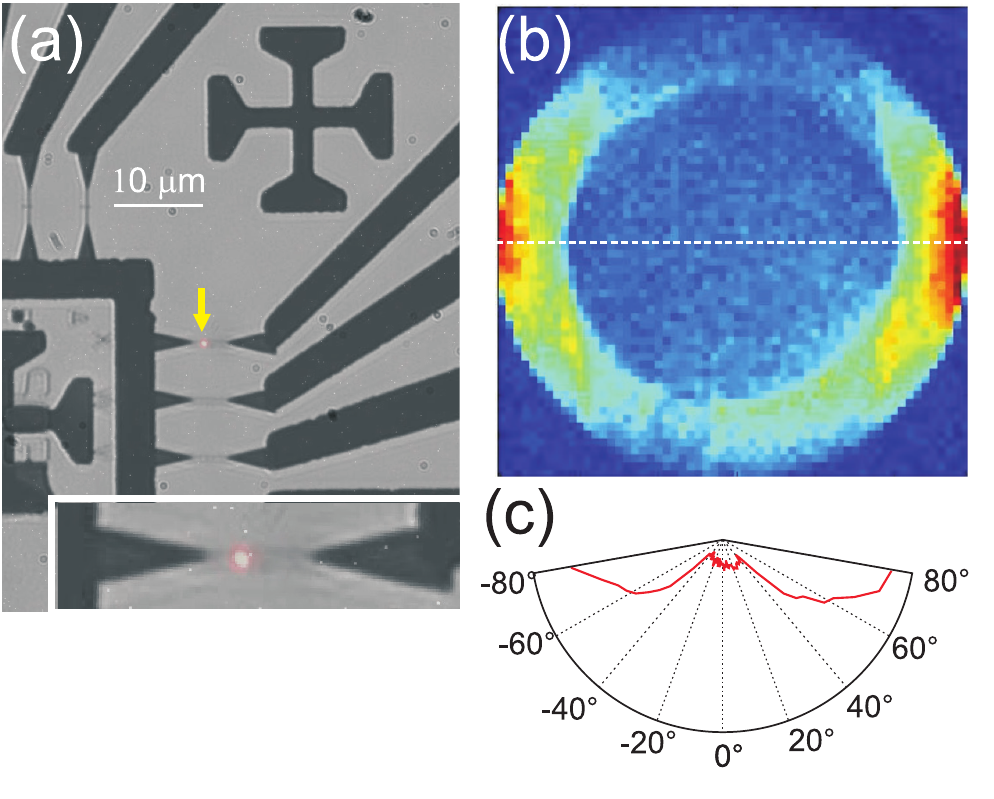}
    \caption{(a) Optical micrograph representing an overlay image of the electrodes system with an image of the light emitted by the antenna (arrow). A magnified view of the optical tunneling gap antenna is shown in the inset. (b) Fourier plane image representing the projected angular distribution of the light emission. (c) Polar plot of the emission along the dashed line in (b). Images are in false color.}
  \label{twolobes}
\end{figure}

For smaller nanowires, the electromigration process does not necessarily lead to a junction located along the nanowire itself. Instead, the density of defects causing current-crowding points produces the tunneling feedgap near the source electrode. Figure~\ref{onelobe} (a) and (b) illustrate this for two 1500~nm long nanowires electromigrated with inverted voltage polarities. In both cases, the location of the light-emitting region is off-centered and is located near the source electrode which provides a simple method to fix the antenna feedgap in a desired location. The emission diagram of the photon source is strongly affected by this position asymmetry as displayed in Fig.~\ref{onelobe} (c) and (d) showing the corresponding Fourier planes. The angular distributions are redirected from a two-lobe configuration to a single intense lobe oriented by the nanowire lead with a maximum emission at $\pm$ 60$^{\circ}$. The measured directivities are now 18.27~dB and 18.08~dB, respectively. The figure of merit of the antenna directionality defined by the front-to-back ratios $F/B$ are estimated from the polar plots in Fig.~\ref{onelobe}(e) and (f). At $\pm$ 60$^{\circ}$, we measure $F/B$=10~dB and $F/B$=6.1~dB for the two devices. These values compare very well to those measured from multi-element designs such the Yagi-Uda geometry~\cite{curto10} and log-periodic optical antennas~\cite{Pavlov2012}.

\begin{figure}
\includegraphics[width=9cm]{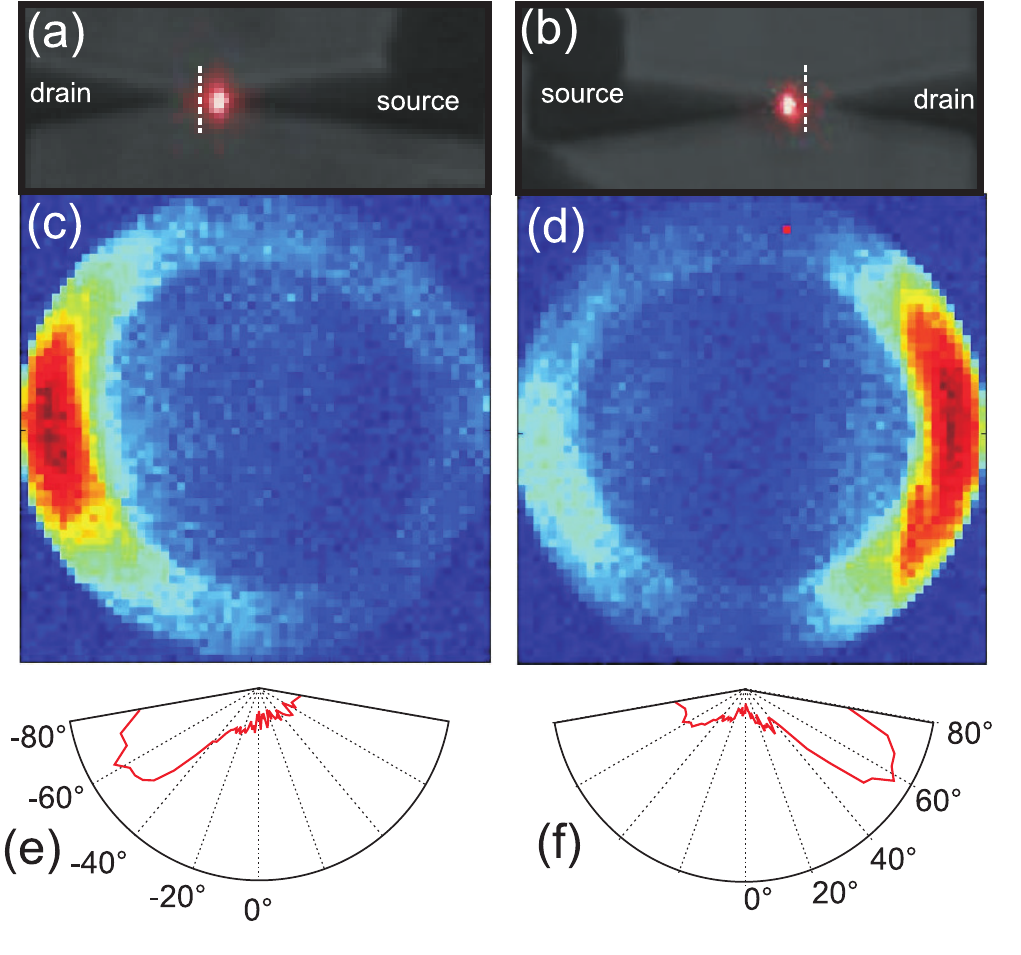}
    \caption{(a) and (b) are false color images of two different light emitting electron pumped antennas contacted by 1500~nm long nanowire. By choosing the polarity during the electromigration, the position the junction can be off-centered towards the source of electrons. The dashed lines are the approximative centers of the nanowires. (c) and (d) are corresponding Fourier planes showing a single emission lobe oriented towards the nanowire. (e) and (f) are polar plots of the emission diagram along the direction of maximum emission.}
  \label{onelobe}
\end{figure}

To summarize, we introduce a new paradigm for optical antennas by developing directive electron-fed light-emitting devices acting as a nanoscale transducer of electrical power. Upon injecting electrons in the feedgap of the antenna, an unconventional emission spectrum is recorded whereby the electromagnetic energy of the emitted photons exceeds the energy of the electrons. We interpret this overbias light emission by the spontaneous emission of a hot electron distribution. The mechanism is as follows: the voltage drop $V_{\rm bias}$ occurs mostly on the antenna gap that is much smaller than the electron-phonon energy exchange length $l_{\rm energy}$. Electrons are accelerated by the electric field to form a hot distribution within a region of a few tens of nanometers near the antenna feedgap.  Because the hot electrons do not efficiently exchange energy with the phonons, the thermalization of the distribution occurs by electron collisions with the antenna borders and spontaneously emits a black body radiation corresponding to an electron temperature up to 2000~K. Thus, two conditions must be satisfied by the optical antennas to emit light from a hot electron gas. First, the voltage drop responsible for pumping the  electron sub-system must be on a length scale smaller than $l_{\rm energy}$, and second, hot electrons must collide with some obstacles to generate an electron's spontaneous emission. Both conditions are met in the antenna gap. 
Our approach suggests that all-metal optical antennas can be integrated as an interface device between an electronic layer and a photonic layer. Of importance for such a device is to what extend the antenna can be electrically modulated. Thermal processes are usually plagued by slow dynamics. However, the emission characteristics of the antennas reported here is dictated by the relaxation dynamics of the hot electrons, which can be as fast as a few ps in metal nanostructures~\cite{Gamaly11}.

The authors declare no competing financial interest.

\begin{acknowledgement}

The research leading to these results has received fundings from the European Research Council under the European Community's Seventh Framework Program FP7/2007--2013 Grant Agreement no 306772,  the Labex ACTION (contract ANR-11-LABX-01-01) and the regional program PARI Photcom. A.U. and I. P.  acknowledge the support of the Russian Foundation for Basic Research under Project
No. 13-08-01438.  

\end{acknowledgement}

\begin{suppinfo}

%This will usually read something like: ``Experimental procedures and
%characterization data for all new compounds. The class will
%automatically add a sentence pointing to the information on-line:
\section{Experimental setup}

Figure~\ref{setup} depicts a simplified sketch of the system used to characterize electrically and optically the electron-fed optical antennas discussed in the main section of the manuscript. The system is built from an inverted optical microscope (Nikon Eclipse) equipped with an oil immersion objective. Two charge-coupled device (CCD) cameras are placed at the different the exit ports of the microscope. A first CCD (Andor, Luca) records a plane conjugate of the object plane ($\Pi^{'''}$). A second CCD (Andor, Ikon) records a plane conjugate to the Fourier plane of the microscope ($\Sigma^{'}$) to evaluate the angular distribution of the emitted photons. To spectrally decomposed the light, we use a spectrograph (Andor, Shamrock) positioned at ($\Pi^{''}$). The electrical activation and characterization is performed by soldering copper leads to a set of macroscopic gold electrodes individually contacting the nanowires. The direct-current (DC) voltage bias $V_{\rm bias}$ is provided by a control electronic (RHK tech, R9). The differential conductance of the nanowire is constantly monitored during and after the electromigration by superposing a small sinusoidal alternative bias to the DC bias. The modulated current contribution is extracted by a lock-in amplifier (Zurich Instruments, HF2LI). 

\begin{figure}
\includegraphics[width=9cm]{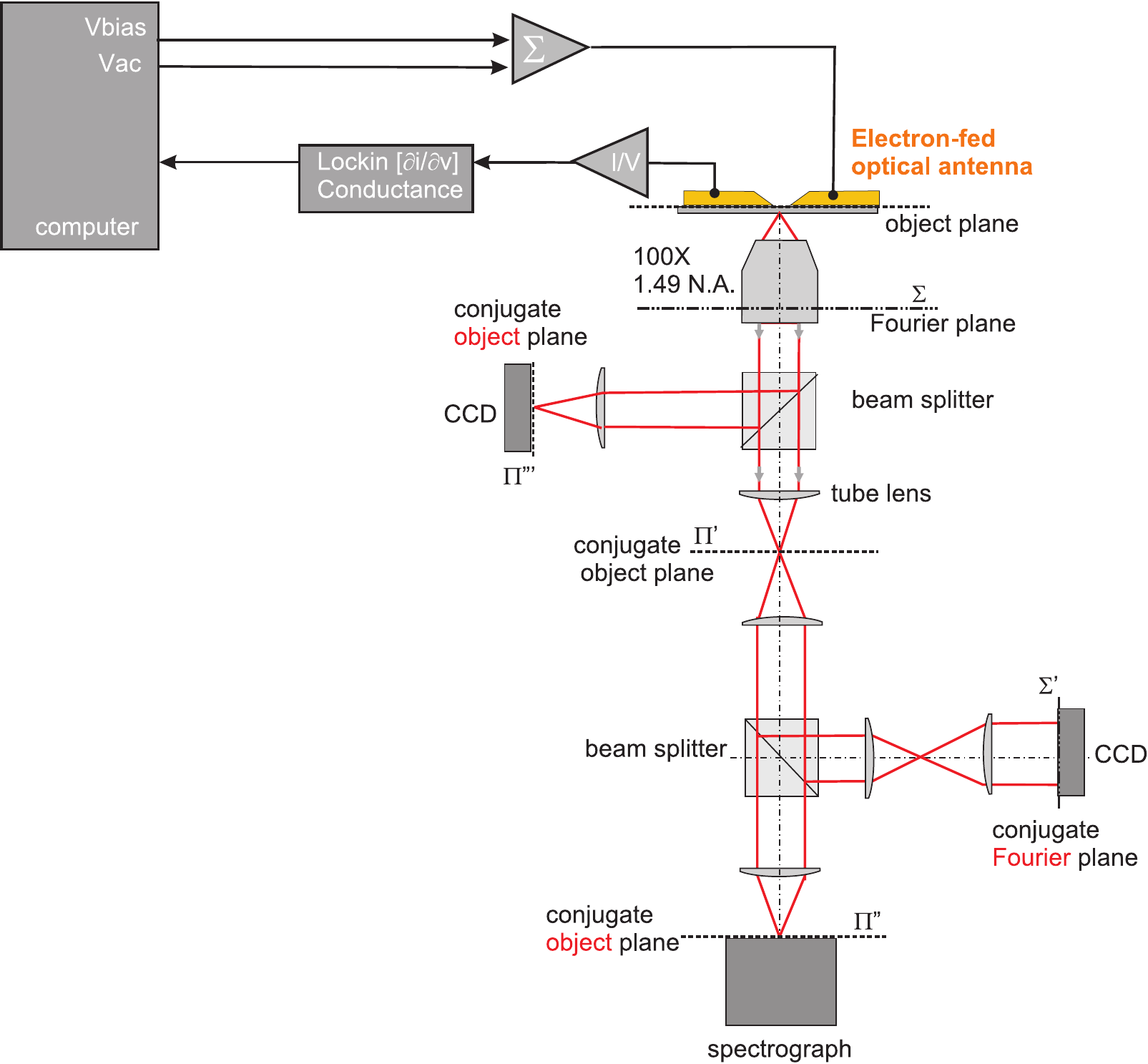}
    \caption{Description of the experimental apparatus used to excite/measure the electrical characteristics of the junction and collect/analyze the emitted photons.}
  \label{setup}
\end{figure}

\section{One-dimensional model for the injected power into a hot-electron system}.

Figure~\ref{1Dmodel} schematically describes the one-dimensional problem. Tunneling electrons are injected into a Au nanowire at $z$=0. The nanowire is contacted to a bus-bar electrode at $z=L_{\rm nw}$ which is also a drain for transporting heat away from the injection region.

\begin{figure}
\includegraphics[width=9cm]{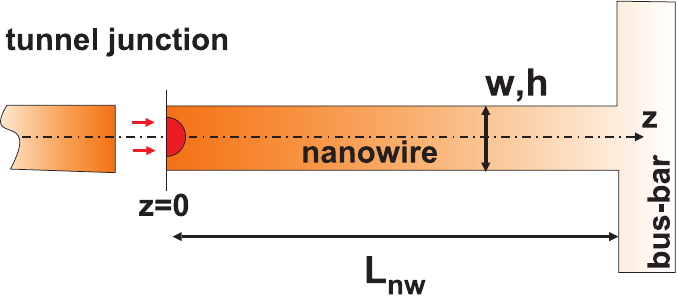}
    \caption{Description of the one-dimensional model. A nanowire with a length $L_{\rm nw}$ and a rectangular section $A=h\times w$ is electrically connected to a bus-bar electrode. Tunneling electrons are injected into the nanowire at $z$=0.}
  \label{1Dmodel}
\end{figure}

The electrical power injected into the nanowire near $z$=0 is $P= I_{\rm T} V_{\rm bias}$ where $I_{\rm T}$ and $V_{\rm bias}$ are the tunneling current and the bias applied across the tunnel junction, respectively. For long nanowire $L_{\rm nw}>>h,w$ where $h$ is the height of the gold nanowire and  $w$ the width, the stationary temperature distribution can be found from the stationary one-dimensional heat equation:

\begin{equation}
\frac{\partial}{\partial z}\left(\kappa(T_e)\frac{\partial T_e}{\partial z}\right)-W_{e\rightarrow L}(T_e, T_L)+p(z)=0
\label{eq2}
\end{equation}
where $T_e$ is the electron temperature, $\kappa = b\times T_e$ is the electron thermal conductivity, $p(z)$ is the distribution of the injected power along the nanowire. The term $W_{e\rightarrow L}(T_e, T_L)$ describes cooling of the electrons to the lattice at temperature $T_L$.

Equation~\ref{eq2} must be solved using the boundary condition  $T_e(z=L_{\rm nw})=T_0$, where $T_0$ is the temperature of the heat drain to which the nanowire is connected. We assume in the following that $T_0=T_L$. For modeling, one can assume that the power is injected just at $z$=0, that is $p(z)  \equiv \delta(z)$.  In this case, instead of Eq.~\ref{eq2}, we can solve
\begin{equation}
\frac{\partial}{\partial z}\left(b\times T_e\frac{\partial T_e}{\partial z}\right)-W_{e\rightarrow L}(T_e, T_L)=0
\label{eq3}
\end{equation}

with the boundary conditions

\begin{eqnarray}
-b\times T_e\frac{\partial T_e}{\partial z} \bigg\vert_{z=0}&=&\frac{P}{A}\\
T_e(z=L_{\rm nw})&=&T_0
\label{eq4}
\end{eqnarray}

In the bulk, electrons cooling to the lattice occurs by generating acoustic bulk phonons, and in this case one writes the term $W_{e\rightarrow L}(T_e, T_L)=H \times (T_e-T_L)$. But in a system with a transverse size $L$ shorter than the electron energy loss length $l_{\rm energy}$ for electron-bulk phonons interaction as in the electromigrated feedgap discussed here, the collisions of electrons with the walls of system become more important. Following the paper by Tomchuk and Fedorovich~\cite{Federovich00,Tomchuk66}, the energy transferred to the lattice writes:
\begin{equation}
W_{e\rightarrow L}(T_e, T_L)=H' \times (T^2_e-T^2_L)
\label{eq5}
\end{equation}

with 
\begin{equation}
H'=\frac{\pi^2}{4}\frac{
k_B^2m^2}{M\hbar^3}E_F\frac{1}{ L}
\label{eq6}
\end{equation}

Here $m$ and $M$ are the electron and atomic masses, respectively, $k_B$ is the Boltzmann constant, $E_F$ is the Fermi energy and $L$ is the distance in which electrons are colliding with the surface. This distance depends on the ill-defined junction geometry resulting from the electromigration process and is thus proportional to the square root of an effective area: $L\sim\sqrt{A_{\rm eff}}$. 
Equation~\ref{eq3} now writes
\begin{equation}
\frac{\partial}{\partial z}\left(b\times T_e\frac{\partial T_e}{\partial z}\right)-H'\times(T^2_e-T^2_L)=0
\label{eq7}
\end{equation}
with the solution 

\begin{equation}
T_e=T_L\sqrt{1+\frac{2P}{A_{\rm eff}}\sqrt{\frac{1}{2bH'T^4_L}}\frac{\exp(-z/z_0)-\exp(-2L_{\rm nw}/z_0)\exp(-z/z_0)}{1+\exp(-2L_{\rm nw}/z_0)}}\label{eq8}
\end{equation}

where $z_0=\sqrt{b/2H'}$ is the heated length of the nanowire. Since the maximum temperature is reached at $z=0$ and assuming that the length of the nanowire is much larger than its heated length, i.e. $L_{\rm nw}>>z_0$, the expression Eq.~\ref{eq8} reduces to

\begin{eqnarray}
T_e&=&T_L\sqrt{1+\frac{2P}{A_{\rm eff}}\sqrt{\frac{1}{2bH'T^4_L}}\frac{\exp(1-\exp(-2L_{\rm nw}/z_0)}{1+\exp(-2L_{\rm nw}/z_0)}}\\
&=&T_L\sqrt{1+\frac{2P}{A_{\rm eff}}\sqrt{\frac{1}{2bH'T^4_L}}}\\
&=&T_L\sqrt{1+\frac{\alpha'}{(k_BT_L)^2}I_{\rm T} V_{\rm bias}}
\label{eq9}
\end{eqnarray}

In form, Eq.~\ref{eq9} coincides with the result by Tomchuk-Fedorovich~\cite{Federovich00,Tomchuk66}  for nanoparticle of size $L$. The term $\alpha'$ depends on the characteristics of the interaction between electrons and the lattice as well as the coefficient $b$ relating the heat conductivity and the electronic temperature. 

\begin{equation}
\alpha'=\frac{2k_B^2}{A_{\rm eff}\sqrt{2bH'}}=\frac{4k_B}{L^{3/2}m\pi\sqrt{2b\frac{E_F}{M\hbar^3}}}
\label{eq10}
\end{equation}

The electronic temperature given by Eq.\ref{eq9} can be recasted in the form:
\begin{equation}
k_BT_e=\sqrt{(k_B T_L)^2+\alpha' I_{\rm T} V_{\rm bias} }
\label{eq11}
\end{equation}
and if $T_e>>T_L$, Eq~\ref{eq11} reduces to:

\begin{equation}
k_BT_e=\sqrt{\alpha' I_{\rm T} V_{\rm bias} }
\label{eq11prime}
\end{equation}

\section{Comparison with the electronic temperature of a constriction deduced from noise measurement}.

In the work by Chen \emph{et al.}~\cite{NatelsonSR14}, the thermal conductance of the electrons writes:

\begin{equation}
C_e=\kappa\frac{A}{l}
\label{eq12}
\end{equation}

where $\kappa$ is the thermal conductivity, $A$ the constriction area and $l$ its length. Feeding the Wiedmann-Franz law in Eq.~\ref{eq12} ($\kappa=L_{\rm Lorenz}T\sigma $), 

\begin{equation}
C_e= L_{\rm Lorenz}T\sigma\frac{A}{l}=L_{\rm Lorenz} TG
\label{eq13}
\end{equation}
where $L_{\rm Lorenz} $ is the Lorenz number ($L_{\rm Lorenz} =\pi^2 k_B^2/3e^2$), $\sigma$ the electrical conductivity, $G$ the electrical conductance of the constriction, and $T$ the average temperature of the system. In correspondence with Eq.~\ref{eq13}, the thermal power delivered by the heated constriction is
\begin{equation}
P_{\rm out}=C_e\times(T-T_0)=(L_{\rm Lorenz}TG)\times(T-T_0)=\frac{T+T_0}{2}L_{\rm Lorenz} G\times(T-T_0)= \frac{1}{2}L_{\rm Lorenz}G\times(T^2-T_0^2)
\label{eq14}
\end{equation}

Equation~\ref{eq14} is expression established by Chen and co-workers~\cite{NatelsonSR14}.  The power dissipated in the constriction is $P_{\rm in}=\alpha^{\rm fraction}GV_{\rm bias}^2$ where $\alpha^{\rm fraction}$ is the fraction of Joules heating dissipated inside the constriction, reported at 2\%.  Balancing Eq.~\ref{eq14} with $P_{\rm in}$ and solving for the $T$ leads to the following expression: 
\begin{equation}
T=\sqrt{T_0^2+\frac{2\alpha^{\rm fraction}V_{\rm bias}^2}{L_{\rm Lorenz}}}=T_0\sqrt{1+\frac{2\alpha^{\rm fraction}P}{T_0^2GL_{\rm Lorenz}}}
\label{eq15}
\end{equation}

If $2\alpha^{\rm fraction}P/T_0^2GL_{\rm Lorenz}>>1$, the temperature of the system writes 

\begin{equation}
T=\sqrt{\frac{2\alpha^{\rm fraction}P}{GL_{\rm Lorenz}}}=\sqrt{\frac{2\alpha^{\rm fraction}Pk_B^2}{GL_{\rm Lorenz}k_B^2}}=\sqrt{\frac{\alpha''P}{k_B^2}}
\label{eq16}
\end{equation}

or equivalently 
\begin{equation}
k_BT=\sqrt{\alpha''P}
\label{eq18}
\end{equation}

where $\alpha''=2\alpha^{\rm fraction}k_B^2/GL$. We note here that both in Chen's work~\cite{NatelsonSR14} and in our work, the temperature of the system depends on the square root of the electrical power fed into the system (Eq.~\ref{eq11prime}). Assuming a ballistic constriction characterized by a single conductance channel, i.e $G=G_0=2e^2/h$, and with $L_{\rm Lorenz}=\pi^2 k_B^2/3e^2$

\begin{equation}
\alpha''=\frac{6}{\pi}\alpha^{\rm fraction}\hbar
\label{eq19}
\end{equation}

Using the reported value of $\alpha^{\rm fraction}=2\%$, $\alpha''=0.04$ in unit of $\hbar$. 
Let us compare this quantity with the experimental value of our work using the one-dimensional formalism described in the previous section. Like in the work by Chen (Eq.~\ref{eq18}), the electronic temperature depends  on the $\sqrt{\alpha' I_{\rm T}V_{\rm bias}}$ with $\alpha'$ given by expression in Eq.~\ref{eq10}: 
\begin{equation}
\alpha'=\frac{4k_B}{L^{3/2}m\pi\sqrt{2b\frac{E_F}{M\hbar^3}}}=\frac{4k_B}{L^{3/2}m\pi\sqrt{\frac{2bE_F}{M\hbar}}}\hbar
\label{eq20}
\end{equation}

Feeding $L=$33~nm deduced from the experimentally inferred $\alpha'$ and the published value of $b=0.03$ for a nanowire with a diameter of 1~nm~\cite{kelly08}, we find $\alpha'=0.078\hbar$ which is twice the value derived from the work of Chen \emph{et al.} \cite{NatelsonSR14}. This two-fold difference is in agreement with the reported electronic temperatures deduced from the experimental data in both set of experiments.

\end{suppinfo}

%\bibliography{/Users/Alex/Documents/BIBLIOGRAPHY/references}
\providecommand*\mcitethebibliography{\thebibliography}
\csname @ifundefined\endcsname{endmcitethebibliography}
  {\let\endmcitethebibliography\endthebibliography}{}

\end{document}